\documentclass{iauc}
\usepackage{graphicx}
\pubyear{2005}
\volume{199}
\pagerange{1--}
\jname{Probing Galaxies through Quasar Absorption Lines}
\editors{P. R. Williams, C. Shu, and B. M\'{e}nard, eds.}

\newcommand{\cwcHI}{\hbox{{\rm H}\kern 0.1em{\sc i}}}
\newcommand{\cwcHII}{\hbox{{\rm H}\kern 0.1em{\sc ii}}}
\newcommand{\cwcLya}{\hbox{{\rm Ly}\kern 0.1em$\alpha$}}
\newcommand{\cwcLyb}{\hbox{{\rm Ly}\kern 0.1em$\beta$}}
\newcommand{\cwcMgI}{\hbox{{\rm Mg}\kern 0.1em{\sc i}}}
\newcommand{\cwcMgII}{\hbox{{\rm Mg}\kern 0.1em{\sc ii}}}
\newcommand{\cwcFeII}{\hbox{{\rm Fe}\kern 0.1em{\sc ii}}}
\newcommand{\cwcCII}{\hbox{{\rm C}\kern 0.1em{\sc ii}}}
\newcommand{\cwcCIII}{\hbox{{\rm C}\kern 0.1em{\sc iii}}}
\newcommand{\cwcCIV}{\hbox{{\rm C}\kern 0.1em{\sc iv}}}
\newcommand{\cwcNV}{\hbox{{\rm N}\kern 0.1em{\sc v}}}
\newcommand{\cwcOVI}{\hbox{{\rm O}\kern 0.1em{\sc vi}}}
\newcommand{\cwcSiIV}{{\rm Si}\kern 0.1em{\sc iv}}
\newcommand{\cwcSiII}{\hbox{{\rm Si}\kern 0.1em{\sc ii}}}
\newcommand{\cwcNII}{\hbox{{\rm N}\kern 0.1em{\sc ii}}}

\title[MgII Absorption Galaxies] 
{MgII Absorption through Intermediate Redshift Galaxies}

\author[Churchill, Kacprzak, \& Steidel]   
{Christopher W. Churchill$^1$,
Glenn G. Kacprzak$^1$ 
\break \and Charles C. Steidel$^2$}

\affiliation{$^1$Department of Astronomy, New Mexico State University, \break 
Las Cruces, NM 88003, USA \break 
email: cwc@nmsu.edu, glennk@nmsu.edu \\[\affilskip]
$^2$Department of Astronomy, California Institute of Technology, \break 
Pasadena, CA 91125, USA \break 
email: ccs@astro.caltech.edu}

\begin{document}

\maketitle

\begin{abstract}

The current status and remaining questions of {\cwcMgII} absorbers are
reviewed with an eye toward new results incorporating high quality
{\it Hubble Space Telescope\/} images of the absorbing galaxies.  In
the end, we find that our current picture of extended gaseous regions
around galaxies at earlier epochs is in need of some revision; {\cwcMgII}
absorbing ``halos'' appear to be patchier and their geometry less
regular than previously inferred.  We also find that the so-called
``weak'' {\cwcMgII} absorbers are associated with normal galaxies over a
wide range of impact parameters, suggesting that this class of
absorber does not strictly select low surface brightness, dwarf
galaxies, or IGM material.  We emphasize the need for a complete
survey of the galaxies in quasar fields, and the importance of
obtaining rotation curves of confirmed absorbing galaxies.


\end{abstract}

\firstsection 

\section{Introduction}

Studying {\cwcMgII} absorption lines holds the promise of placing
powerful constraints on scenarios of galactic formation and evolution.
{\cwcMgII} absorption is arguably the best tracer of metal--enriched
gas {\it associated with galaxies\/}.  Magnesium is an
$\alpha$--process element ejected by supernovae into interstellar,
intergroup, and intergalactic space from the time of the very first
generation of stars to the present epoch.  Furthermore, {\cwcMgII}
absorption arises in gas spanning more than five decades of neutral
hydrogen column density, from $\log N({\cwcHI}) \simeq 15.5$ to
greater than $20.5$~[atoms~cm$^{-2}$] (e.g., \cite[Bergeron \&
Stasi\'{n}ska 1986]{cwcref:bs86}; \cite[Steidel \& Sargent
1992]{cwcref:ss92}; \cite[Churchill {\etal} 2000a]{cwcref:archiveI});
it therefore probes a wide range of {\cwcHI} environments associated
with star formation, i.e., galaxies.

\begin{figure}
\includegraphics[width=5.3in,angle=0]{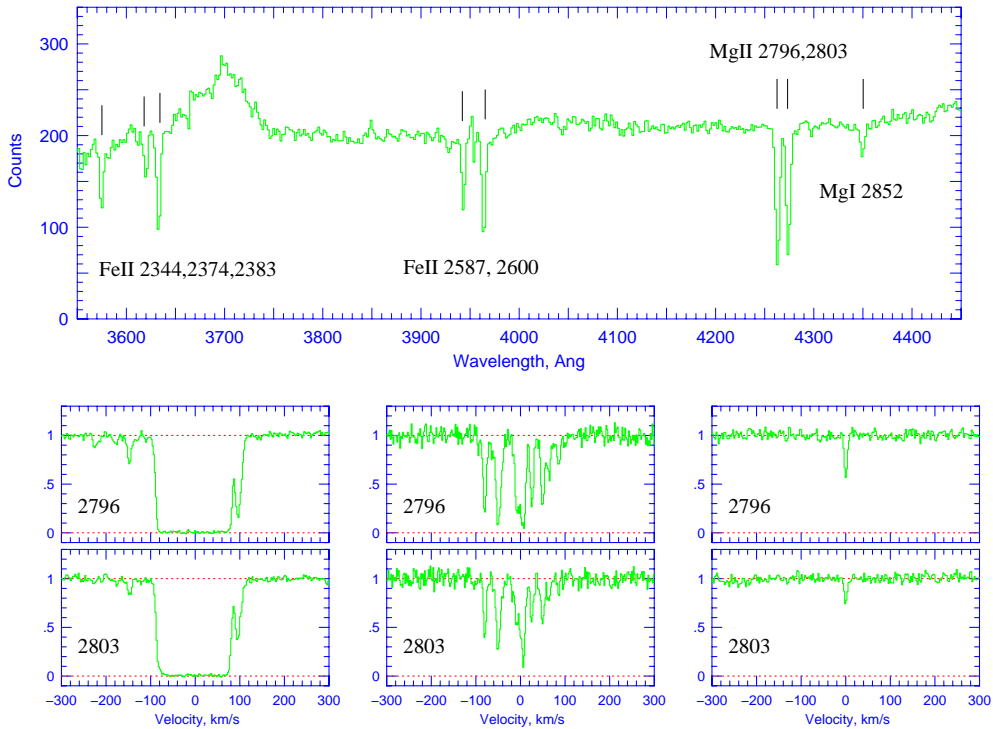}
\caption{--- (upper) A typical strong {\cwcMgII} absorber at a
resolution of $1.5$~{\AA} showing the {\cwcMgII} doublet and
accompanying {\cwcMgI} and {\cwcFeII} absorption (from the survey of
Steidel \& Sargent 1992).  --- (lower) The kinematic ``morphology'' of
three classes of {\cwcMgII} absorbers, (left) DLA/{\cwcHI}--rich,
(center) LLS/classic, and (right) weak system (from the survey of
Churchill \& Vogt 2001).}
\label{cwcfig:types}
\end{figure}

~{\cwcMgII} systems are observed in the optical over the large
redshift range $0.3\leq z\leq 2.2$, are common (about one for every
unit of redshift), and are easily identified in quasar spectra.  Shown
in Figure~\ref{cwcfig:types} (upper panel) is an example of a typical
{\cwcMgII} system and its associated {\cwcMgI} and {\cwcFeII}
absorption.  The tall--tale signature of these systems is unambiguous.
In order to exploit these systems for studies of galaxy evolution, it
is imperative to establish the connections between the absorbing gas
properties and the galaxy properties.

The kinematics of {\cwcMgII} absorbers loosely correspond to the type
of galaxy environment and {\cwcHI} column density regime being probed.
In Figure~\ref{cwcfig:types} (lower panels), three ``classes'' of
absorber kinematics are presented.  These classes are more generally
based upon a multivariate statistical analysis incorporating
associated high ionization and {\cwcHI} absorption properties
(\cite[Churchill {\etal} 2000b]{cwcref:archiveII}).

\noindent \underbar{DLA/{\cwcHI}-Rich Systems} (lower left panel):
When the {\cwcMgII} profiles are fully saturated with no discernible
velocity substructure, the system is often a damped Ly~$\alpha$
absorber (DLA) with $\log N({\cwcHI}) \geq 20.3$ cm$^{-2}$. The
{\cwcMgII} kinematics range between 100--200 km~s$^{-1}$ and are
rarely seen to have ``kinematic subsystems'' (higher velocity
absorption).  The associated galaxies are observed to be a
``mixed--bag'' of morphologies and luminosities, including low surface
brightness galaxies (\cite[Bowen, Tripp, \& Jenkins
2001]{cwcref:bowen01}) and sub--$L^{\ast}$ to $L^{\ast}$ galaxies
(\cite[Le Brun {\etal} 1997; Rao {\etal} 2003]{cwcref:lebrun97,rao03};
also see Figure~\ref{cwcfig:impacstr}).

\noindent \underbar{Classic Systems} (lower center panel): When the
{\cwcMgII} profiles exhibit a complex run of optical depth with
velocity, the system is often a Lyman limit system (LLS) with $\log
N({\cwcHI}) \geq 17.3$ cm$^{-2}$.  The {\cwcMgII} kinematics range
between 50--400 km~s$^{-1}$ and often have higher velocity weak,
narrow, kinematic subsystems.  The kinematics can vary greatly from
system to system and it is difficult to identify a ``poster--child''
example. The associated galaxies are often fairly normal appearing,
bright, spiral galaxies (\cite[Steidel {\etal} 1998]{steidel98}; also
see Figure~\ref{cwcfig:impacstr}).

\noindent \underbar{Weak Systems} (lower right panel): At very low
equivalent width, $W_r \simeq 0.1$~{\AA}, the {\cwcMgII} profiles are
often unresolved in high resolution spectra. The system is often a
sub--LLS absorber with $\log N({\cwcHI}) \leq 16.8$ cm$^{-2}$.  Some
weak absorbers have multiple ``clouds'' spread over 10--100
km~s$^{-1}$, but most are single clouds.  It has been suggested that
these systems are associated with low surface brightness galaxies
(\cite[Churchill \& Le~Brun 1997]{cwcref:cl98}), intergalactic star
forming ``pockets'' (\cite[Rigby {\etal} 2002]{cwcref:rigby02}) or
dwarf galaxies (\cite[Zonak {\etal} 2004]{cwcref:zonak04}).

\section{{\cwcMgII} Absorbers: A Brief Overview}

There is a rich history of the study of {\cwcMgII} absorbers that
cannot be given full justice in this short paper.  However, a
non--exhaustive survey of the literature provides a reasonable
representation of the variety of excellent contributions over the last
decade and a half.  In Table~\ref{cwctab:mgiirefs}, a brief list of
{\cwcMgII} absorber related literature is presented by three
categories: spectroscopic surveys, studies of the galaxy properties,
and modeling of galaxy halos and ionization conditions.

\begin{table}[bh]
\begin{center}
\caption{Non-Exhaustive {\cwcMgII} Absorber Literature}
\label{cwctab:mgiirefs}
\begin{tabular}{lll}\hline
Spectroscopic                  & Galaxy Connection                 & Modeling \\\hline
\underbar{\bf UV Surveys}      & \underbar{\bf Surveys}            & \underbar{\bf Galaxy Halos} \\
\cite{cwcref:archiveI}         & \cite{cwcref:yanny90}             & \cite{cwcref:gruenwald93} \\
\cite{cwcref:cwc01}            & \cite{cwcref:lb90}                & \cite{cwcref:srianand93} \\
\underbar{\bf Optical Surveys} & \cite{cwcref:bb91}                & \cite{cwcref:phillips93} \\
\cite{cwcref:ltw87}            & \cite{cwcref:bechtold92}          & \cite{cwcref:mo96} \\
\cite{cwcref:tytler87}         & \cite{cwcref:lebrun93}            & \cite{cwcref:cc96} \\ 
\cite{cwcref:ssb88b}           & \cite{cwcref:drinkwater93}        & \cite{cwcref:cc98} \\ 
\cite{cwcref:caulet89}         & \cite{cwcref:steidel93}           & \cite{cwcref:lin01} \\
\cite{cwcref:pb90}             & \cite{cwcref:sdp94}               & \cite{cwcref:bond01b}  \\
\cite{cwcref:biosse92}         & \cite{cwcref:bbp95}               & \underbar{\bf Ionization Conditions} \\
\cite{cwcref:ss92}             & \cite{cwcref:steidel95}           & \cite{cwcref:bs86} \\
\cite{cwcref:aldcroft94}       & \cite{cwcref:gb97}                & \cite{cwcref:bergeron94} \\
\cite{cwcref:malhotra97}       & \cite{cwcref:bouche04}            & \cite{cwcref:dk95} \\
\cite{cwcref:weakI}            & \cite{cwcref:menard05}            & \cite{cwcref:cl98} \\
\cite{cwcref:rt00}             & \underbar{\bf Gas Kinematics}     & \cite{cwcref:cvc03} \\
\cite{cwcref:cv01}             & \cite{cwcref:lb92}                & \underbar{\bf Multiphase Kinematics} \\
\cite{cwcref:cvc03}            & \cite{cwcref:csv96}               & \cite{cwcref:cc99} \\ 
\cite{cwcref:ellison04}        & \cite{cwcref:archive99}           & \cite{cwcref:rigby02} \\ 
\cite{cwcref:nestor05}         & \cite{cwcref:archiveII}           & \cite{cwcref:charlton03} \\
\cite{cwcref:prochter05}       & \underbar{\bf Morphologies}       & \cite{cwcref:ding03} \\ 
\underbar{\bf IR Studies}      & \cite{cwcref:bbp96}               & \cite{cwcref:zonak04} \\
\cite{cwcref:elston96}         & \cite{cwcref:steidel97}           & \cite{cwcref:masiero05} \\
\cite{cwcref:naoto02}          & \cite{cwcref:steidel98}           & \cite{cwcref:ding05} \\
                               & \cite{cwcref:csk05}               & \\
                               & \cite{cwcref:kcs05}               & \\
                               & \underbar{\bf Gas+Gal Kinematics} & \\
                               & \cite{cwcref:steidel02}           & \\
                               & \cite{cwcref:ellison03}           &  \\\hline
\end{tabular}
\end{center}
\end{table}

\subsection{Spectroscopic Surveys}

The spectroscopic surveys are conveniently sub--divided into the
ultraviolet (UV), optical, and infrared (IR) bands, since these probe
distinct redshift regimes.

The UV surveys include an unbiased low redshift ($z \leq 0.1$)
{\cwcMgII} survey (\cite[Churchill 2001]{cwcref:cwc01}) and a study of
the far--UV high ionization transitions (i.e., {\cwcCIV}, {\cwcNV},
{\cwcOVI}, etc.) and neutral hydrogen transitions (i.e., {\cwcLya},
{\cwcLyb}, etc.) associated with intermediate redshift {\cwcMgII}
absorbers using the FOS spectrograph on board {\it HST\/}
(\cite[Churchill {\etal} 2000a,b)]{cwcref:archiveI,cwcref:archiveII}.

Several optical surveys have yielded a solid statistical picture of
{\cwcMgII} absorbers at intermediate redshifts ($0.3\leq z \leq 2.2$).
In the late 1980s and early 1990s, these surveys involved the slow
methodical accumulation of quasar spectra with resolutions 1--2~{\AA}
and yielded some few hundred absorption systems (\cite[e.g., Lanzetta,
Turnshek, \& Wolfe 1987; Sargent, Steidel, \& Boksenberg 1988b;
Steidel \& Sargent 1992]{cwcref:ltw87,cwcref:ssb88b,cwcref:ss92}).
The equivalent width sensitivity of these surveys was typically
$0.3$~{\AA}.  With the advent of the Sloan Digital Sky Survey (SDSS)
mega--database, \cite{cwcref:nestor05} and \cite{cwcref:prochter05}
cataloged several thousand {\cwcMgII} absorbers ripe for further
study.  However, the sensitivity is not uniform and hovers just below
the $1.0$~{\AA} level for the majority of the quasars.

\begin{figure}[hb]
\includegraphics[width=5.3in,angle=0]{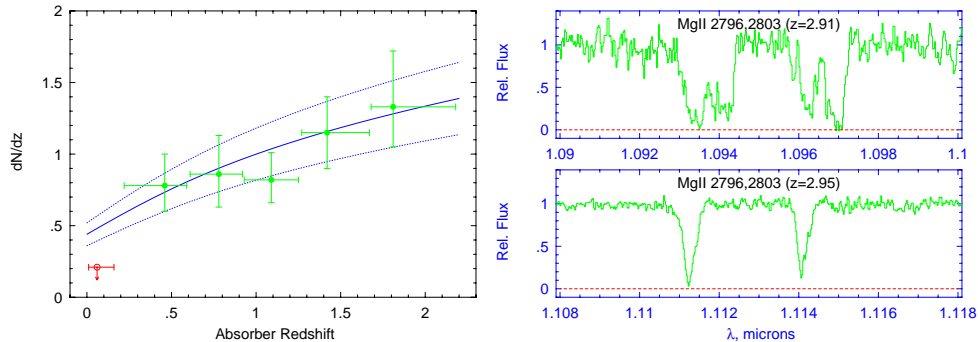}
\caption{--- (left) The redshift path number density of {\cwcMgII}
absorbers with $W_r \geq 0.3$~{\AA} as a function of redshift.  The
solid circles are from Nestor {\etal} (2005), and the open point
(upper limit) is taken from Churchill (2001).  --- (right) Examples of
$z\sim3$ {\cwcMgII} systems discovered in our in--progress survey of
higher redshift {\cwcMgII} absorbers using the IRCS on Subaru.  The
velocity resolution is $\Delta v \simeq 45$~km~s$^{-1}$ in the
absorber rest frame.}
\label{cwcfig:dndz-ircs}
\end{figure}

An important quantity measured from these surveys is the redshift path
density, $dN/dz$.  In Figure~\ref{cwcfig:dndz-ircs} (left panel), the
redshift path density is plotted for SDSS survey of
\cite{cwcref:nestor05} and the FOS {\it HST\/} survey of
\cite{cwcref:cwc01}.  The solid curve is the no--evolution expectation
($\Omega_m=0.3, \Omega_{\Lambda}=0.7$),
\begin{equation}
dN/dz = N(z) (1+z)^2 [\Omega_m (1+z)^3 + \Omega _{\Lambda} ] ^{-1/2} ,
\label{eq:dNdzpar}
\end{equation}
fit to the binned data, where the dotted curves provide the range for
$3~\sigma$ uncertainty in the normalization, $N(z)$.  The $3~\sigma$
upper limit at $z\sim0.05$ (open point) assumes no evolution in the
equivalent width distribution from the intermediate redshift data.

A first glimpse of the {\cwcMgII} absorber kinematics was provided by
\cite{cwcref:pb90} at a resolution of $\Delta v \sim 30$~km~s$^{-1}$.
With the advent of HIRES (\cite[Vogt {\etal} 1994]{cwcref:vogt94}) on
the Keck~I telescope, \cite{cwcref:cv01} were able to resolve the
kinematics of {\cwcMgII} absorption at $\Delta v \sim 6$~km~s$^{-1}$.
These data also allowed \cite{cwcref:weakI} to probe to a sensitivity
of $\sim 0.02$~{\AA}, which uncovered the large population of ``weak''
{\cwcMgII} absorbers.  A low--end cut off in the {\cwcMgII} equivalent
width distribution is yet to be observed.

Charting the properties of {\cwcMgII} absorbers at $z>2.2$ requires IR
spectroscopy.  In the IR, the present status is similar to that of
optical surveys in the late 1980s and early 1990s.  Recently, surveys
of {\cwcMgII} absorbers have become possible using IR spectrographs,
but the work is slow.  In collaboration with Naoto Kobayashi, we have
accumulated $\sim 35$ high redshift quasars for a survey of $2.5\leq z
\leq 4.0$ {\cwcMgII} absorbers.  In Figure~\ref{cwcfig:dndz-ircs}
(right panel), two examples of the $z\sim3$ {\cwcMgII} absorbers are
shown.  We are using NICFPS on the 3.5--meter telescope at the Apache
Point Observatory to image the quasar fields to survey for the
absorbing galaxies at $2.5\leq z \leq 4.0$.

\subsection{The Absorber--Galaxy Connection}

Studies of the galaxies include surveys of the quasar fields, both
targeted and statistical, and targeted studies of the absorbing gas
kinematics, the galaxy morphologies, and the connections between the
galaxy kinematics and the gas kinematics.

The nature of luminous objects associated with {\cwcMgII} absorbers
was an open debate in the early 1990s.  The work of
\cite{cwcref:yanny90} favored a picture in which star forming
sub--galactic fragments at impact parameters of $100$--$200$ kpc gave
rise to {\cwcMgII} absorption.  However, this was at a time when some
scenarios suggested that $z\sim0.2$ was an epoch of active galaxy
formation via sub--galactic fragments.  The seminal work of
\cite{cwcref:bb91} strongly suggested the presence of a bright, $\sim
L^{\ast}$ galaxy within a few tens of kpc from the absorbing gas.
Soon after, \cite[Steidel, Dickinson, \& Persson (1994)]{cwcref:sdp94}
(hereafter, SDP94) firmly established the work of \cite{cwcref:bb91}
and reported the gross properties of the galaxies, including their
rest--frame $B-K$ colors, and $L_B$ and $L_K$ luminosity functions.
\cite{cwcref:steidel95} presented a statistical picture of the
luminosity dependent gas cross sections of {\cwcMgII} absorbing
galaxies that has guided our conventional wisdom to the present.

\begin{figure}[b]
\includegraphics[width=5.3in,angle=0]{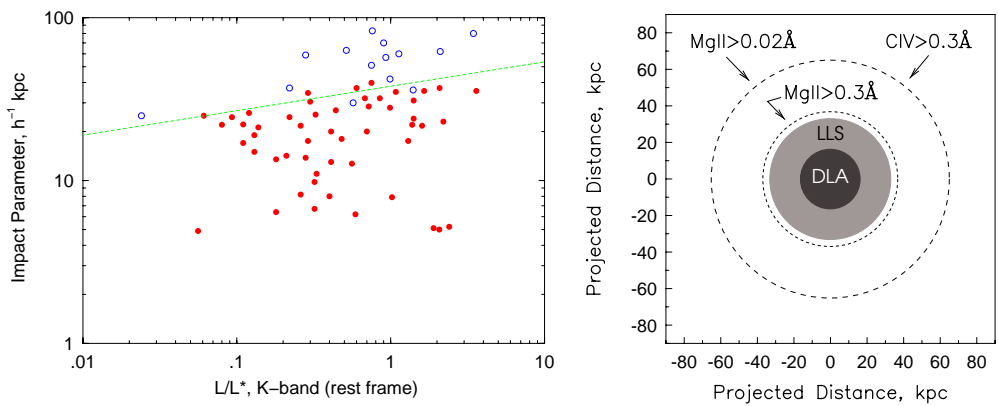}
\caption{ --- (left) Galaxy--quasar impact parameter versus galaxy $K$
luminosity for $W_r \geq 0.3$~{\AA} (adapted from Steidel 1995).  The
solid points are absorbing galaxies and the open points are
non--absorbers taken from ``control'' fields.  The linear relationship
is Eq.~\ref{cwceq:rlk}.  --- (right) The statistical cross section of
various classes of absorbers, including DLAs, LLS, strong and weak
{\cwcMgII} absorbers, and {\cwcCIV} systems.}
\label{cwcfig:rlk-xsec}
\end{figure}

The \cite{cwcref:steidel95} results are worth additional discussion.
In Figure~\ref{cwcfig:rlk-xsec} (left panel) the impact parameters are
plotted versus $K$ luminosity for galaxies selected by $W_r \geq
0.3$~{\AA} {\cwcMgII} absorption.  Absorbing galaxies are plotted as
solid circles and non--absorbing galaxies, based upon ``control''
fields, are plotted as open points.  The dashed line was obtained by
minimizing the number of non--absorbers above the line and minimizing
the number of absorbers below the line.  Using these results,
\cite{cwcref:steidel95} inferred that the ``halo'' size of the
galaxies scales as
\begin{equation}
R(L_K) = 38 h^{-1} \left( L_K/L_K^{\ast} \right) ^{0.15} \quad {\rm kpc} ,
\label{cwceq:rlk}
\end{equation}
that they have near unity covering factor, and that they must be
roughly spherical in shape.  It is otherwise difficult to explain the
lack of non--absorbers below the $R(L)$ relation of
Eq.~\ref{cwceq:rlk}.  Based upon semi--analytical and Monte--Carlo
modeling, \cite{cwcref:lin01} infer that the value of the slope,
$\beta=0.15$, may in fact be steeper due to a limiting magnitude
selection effect in the survey.  We will revisit these inferences in
\S~\ref{cwcsec:rlkdiscuss}.

The redshift path density normalization, $N(z)$, constrains the
product of the number density, $n(z)$, and cross section, $\sigma(z)$,
that is $N(z) = n(z)\sigma(z)$.  Using the $K$ luminosity function of
\cite[SDP94]{cwcref:sdp94} for an estimate for $n(z)$ and applying
Eq.~\ref{cwceq:rlk} for the gas cross section provides a convenient
formalism for the ``halo'' size of an $L^{\ast}$ galaxy,
\begin{equation}
R_{\ast} = 
\bigg[ \frac{\pi c f_c \Phi ^{\ast}}{H_o } 
 \frac{(dX/dz)}{(dN/dz)} \Gamma ( 2\beta -\alpha +1, L_{min}) 
 \bigg] ^{-1/2} ,
\label{eq:rstar}
\end{equation}
where $dN/dz$ is measured, $f_{c}=1$ is the assumed covering factor,
$\Gamma$ is the incomplete Gamma function, $\Phi^{\ast} = 0.03h^3$
Mpc$^{-3}$ and $\alpha =-1$ are the normalization and faint--end slope
of the galaxy luminosity function, $\beta = 0.15$ is the slope of the
$R(L_K)$ relation, $L_{min}=0.05L^{\ast}_K$ is the faint--end cut off
for galaxies giving rise to absorption, and $dX/dz$ is the
differential absorption distance, $dX/dz = (1+z)^2 / \sqrt{\Omega_m
(1+z)^3 +\Omega_{\Lambda}} $, which is proportional to the probability
of intercepting an absorber per unit redshift in a pencil beam survey.

In Figure~\ref{cwcfig:rlk-xsec} (right panel), $R_{\ast}$ is shown for
various regimes of $N({\cwcHI})$ associated with {\cwcMgII} absorbers,
including DLAs, LLS, and strong and weak {\cwcMgII} absorbers.  For
strong {\cwcMgII} absorbers, $R_{\ast} \simeq 40h^{-1}$ kpc and for
weak {\cwcMgII} systems $R_{\ast} \simeq 70h^{-1}$ kpc.  Note that,
statistically, LLS and strong {\cwcMgII} systems appear to be drawn
from the same population.  For comparison, {\cwcCIV} selected absorber
sizes are also shown (\cite[e.g., Sargent, Steidel, \& Boksenberg
1988a]{cwcref:ssb88a}).  \cite{cwcref:nestor05} has performed a more
sophisticated analysis, presenting $R_{\ast}$ as a function of
redshift for a variety of equivalent width cut offs.

A comparison of the gas kinematics and gross properties of the
galaxies was pioneered by \cite{cwcref:lb92}, and later expanded upon
by \cite[Churchill, Steidel, \& Vogt (1996)]{cwcref:csv96}, and
\cite[Churchill {\etal} (1999a,
2000b)]{cwcref:archive99,cwcref:archiveII}.  However, the sample size
remained relatively small, at $\sim 15$ galaxies, and no significant
correlations were found.  Little effort could be applied to
establishing the nature of galaxies associated with weak {\cwcMgII}
absorbers since the galaxies studied were originally selected from
surveys with equivalent width sensitivities of $W_r \geq 0.3$~{\AA}.
However, both the fact that weak {\cwcMgII} absorbers are sub--LLS
systems (\cite[Churchill {\etal} 2000a; Rigby {\etal}
2002]{cwcref:archiveI,cwcref:rigby02}) and the apparent lack of
obvious bright, normal galaxy candidates in the studied fields
(\cite[Churchill {\etal} 1999b]{cwcref:weakI}) allowed for speculation
that weak {\cwcMgII} absorbers were probably associated with LSB
and/or dwarf galaxies at large impact parameters or isolated star
forming pockets in the intergalactic medium (\cite[Churchill \&
Le~Brun 1997; Rigby {\etal} 2002; Zonak {\etal}
2004]{cwcref:cl98,cwcref:rigby02,cwcref:zonak04}).

{\it HST\/} has obtained images of roughly three--dozen quasar fields
with known {\cwcMgII} absorbers.  The images, most of them obtained
using WFPC--2 with the F702W filter, provide a few to several
kiloparsec resolution of intermediate redshift galaxies approximately
at their rest--frame $B$ band. Only a single quasar field, 3C~336, has
been thoroughly studied in that the majority of galaxies within
$50^{\prime\prime}$ of the quasar have been spectroscopically
redshifted (\cite[Steidel {\etal} 1997]{cwcref:steidel97}).

A direct comparison of the galaxy kinematics and the {\cwcMgII}
absorption kinematics has been performed for only six galaxies to date
(\cite[Steidel {\etal} 2002; Ellison, Mall\'{e}n--Ornelas, \& Sawicki
2003]{cwcref:steidel02,cwcref:ellison03}).  The galaxies have been
carefully selected; each a nearly edge--on spiral with the disk
major--axis intersecting the quasar line of sight (though beyond the
visible extent of the disk). In the \cite{cwcref:steidel02} sample,
four of the five galaxies have {\cwcMgII} absorbing gas that traces
the rotation curve.  Simplistic models suggest that the gas ``halo''
corotates with the galaxy, but that the halo rotation lags the galaxy
rotation with height above the disk.  Lagging halos are observed in
local galaxies using 21--cm emission (\cite[Sancisi {\etal}
2001]{cwcref:sancisi01}) and diffuse ionized gas emission
(\cite[Swaters, Sancisi, \& van der Hulst 1997; Rand
2000]{cwcref:swater97,cwcref:rand00}).  In the fifth galaxy, the
absorption is a narrow weak {\cwcMgII} component aligned at the galaxy
systemic velocity. \cite[Ellison {\etal} (2003)]{cwcref:ellison03}
found a different result; some of the {\cwcMgII} absorption is aligned
with the galaxy rotation, but the majority of the gas lies at negative
velocities with respect to the galaxy.  Their result is analogous to
the ``forbidden gas'' observed in 21--cm studies of local galaxies
(\cite[Fraternali {\etal} 2002]{cwcref:fraternali02}). \cite[Ellison
{\etal}]{cwcref:ellison03} favor a scenario in which supperbubbles
give rise to the {\cwcMgII} absorbing gas in the lower halo of the
galaxy (\cite[also see Bond {\etal} 2001a]{cwcref:bond01a}).

This latter type of study holds the greatest promise for understanding
the spatial and kinematic relationship between galaxies and their
extended gaseous components. If we are to place constraints on
scenarios and models of galaxy evolution using quasar absorption
lines, it is imperative that a wide range of galaxy--quasar
line--of--sight orientations are studied in tandem with the galaxy
kinematics.

\subsection{Modeling and Ionization Conditions}

Modeling the geometric structure and ionization conditions of
{\cwcMgII} ``halos'' began in earnest when the connection between
galaxies was first being established observationally (\cite[Gruenwald
\& Viegas 1993; Srianand \& Khare
1993]{cwcref:gruenwald93,cwcref:srianand93}).  Additional
sophistication was applied by \cite{cwcref:mo96}, who developed a
two--phase model in which {\cwcMgII} halos are defined by the cooling
radius of infalling gas.  However, their models predict too few
LLS/{\cwcMgII} absorbers at $z>2$.  \cite{cwcref:cc98} modeled the
kinematic composition of {\cwcMgII} absorbers; they concluded that
disk kinematics likely make a significant contribution to the
absorption profiles.

Photoionization modeling has been applied to observational data by
several teams, following the seminal work of \cite{cwcref:bs86}.
However, it was not until the STIS spectrograph on {\it HST} provided
high resolution UV spectra of absorption kinematics from a wide range
of ionization species, such as {\cwcSiII}, {\cwcSiIV}, {\cwcCII},
{\cwcCIII}, {\cwcCIV}, {\cwcNII}, {\cwcNV}, {\cwcOVI}, and neutral
hydrogen, that multiphase photoionization could be modeled in detail
(\cite[Charlton {\etal} 2003; Ding {\etal} 2003; Zonak {\etal} 2004;
Masiero {\etal} 2005; Ding, Charlton, \& Churchill
2005]{cwcref:charlton03,cwcref:ding03,cwcref:masiero05,cwcref:zonak04,cwcref:ding05}).

From these studies, it is now firmly established that {\cwcMgII}
absorbers are accompanied by a range of multiphase
ionization/kinematic conditions, including separate high ionization
phases in which {\cwcMgII} absorption is absent to $W_r \simeq
0.02$~{\AA}.  Importantly, the neutral hydrogen arises in multiple
phases, highlighting the need for careful application of the Lyman
series and careful separation of the {\cwcHI} contributions to each
phase.

\section{Toward New Insights into {\cwcMgII} Absorbing Galaxies}
\label{cwcsec:rlkdiscuss}

In the early 1990s, intermediate redshift galaxy surveys were in their
incipient stages and {\cwcMgII} absorption galaxy selection was a
clear cut and competitive method for exploring the general evolution
of galaxies to earlier epochs.  One of the first luminosity functions
of $z \sim 0.6$ galaxies was established using {\cwcMgII} absorption
selection (\cite[SDP94]{cwcref:sdp94}).  During this period, it was
always purported that the {\cwcMgII} absorption selection method held
the additional potential of measuring the kinematic, ionization, and
chemical conditions of the gas phases of galaxies.  In terms of direct
case--by--case studies of the galaxies, much of this potential remains
unexplored.

In an effort to build a large database for direct comparisons of the
galaxy morphologies and {\cwcMgII} absorption kinematics, we have
undertaken a program incorporating high spatial resolution {\it HST\/}
images of the quasar fields and HIRES and UVES spectra of the
{\cwcMgII} absorption.  We have been careful to include in our sample
only those absorption--selected galaxies with {\it confirmed\/}
spectroscopic redshifts.  It is important to keep in mind that our
overall picture of {\cwcMgII} absorption selected galaxies has been
based upon a sample of 52 of which 70\% have spectroscopic redshifts
and 30\% are candidates (\cite[SDP94]{cwcref:sdp94}).  By only
including {\it confirmed\/} galaxies, it is our aim to be sure to
illuminate any possible bias that may have influenced what can be
inferred about {\cwcMgII} absorber covering factors, halo sizes and
geometries, and associated galaxy properties (\cite[see Charlton \&
Churchill 1996]{cwcref:cc96}).

The \cite[SDP94]{cwcref:sdp94} survey method was to obtain
spectroscopic redshifts of galaxies in an outward pattern from the
quasar, selecting galaxies with magnitudes and colors consistent with
their being at the absorber redshift.  The method was very effective.
However, it is always possible that galaxies with unexpected
magnitudes and odd colors near the quasars or galaxies at large
separation could have been missed in the survey.  SDP94 also measured
the properties of galaxies in 25 ``control'' quasar fields, which were
selected by the {\it absence\/} of {\cwcMgII} absorbers with $W_r \geq
0.3$~{\AA} in the quasar spectrum.  In this way, they obtained the
properties of ``non--absorbing'' galaxies.  We discuss possible bias
in this approach in \S~\ref{cwcsec:discuss}.

\subsection{Our Sample}

To date, we have 38 galaxies in our sample, 26 of which we also have
HIRES and/or UVES spectra of the absorption.  In some of our fields,
more than one galaxy is at the absorber redshift.  For now, we cannot
be 100\% certain that additional galaxies contribute to the
absorption.  Ultimately, it will be necessary to establish the
redshifts of all galaxies to a fixed absolute magnitude and to a fixed
angular separation from the quasar.  We are exploring the technique of
photometric redshifts in order to place first order constraints on the
redshift distribution of galaxies in the {\it HST\/} fields.

\begin{figure}[hb]
\includegraphics[width=5.3in,angle=0]{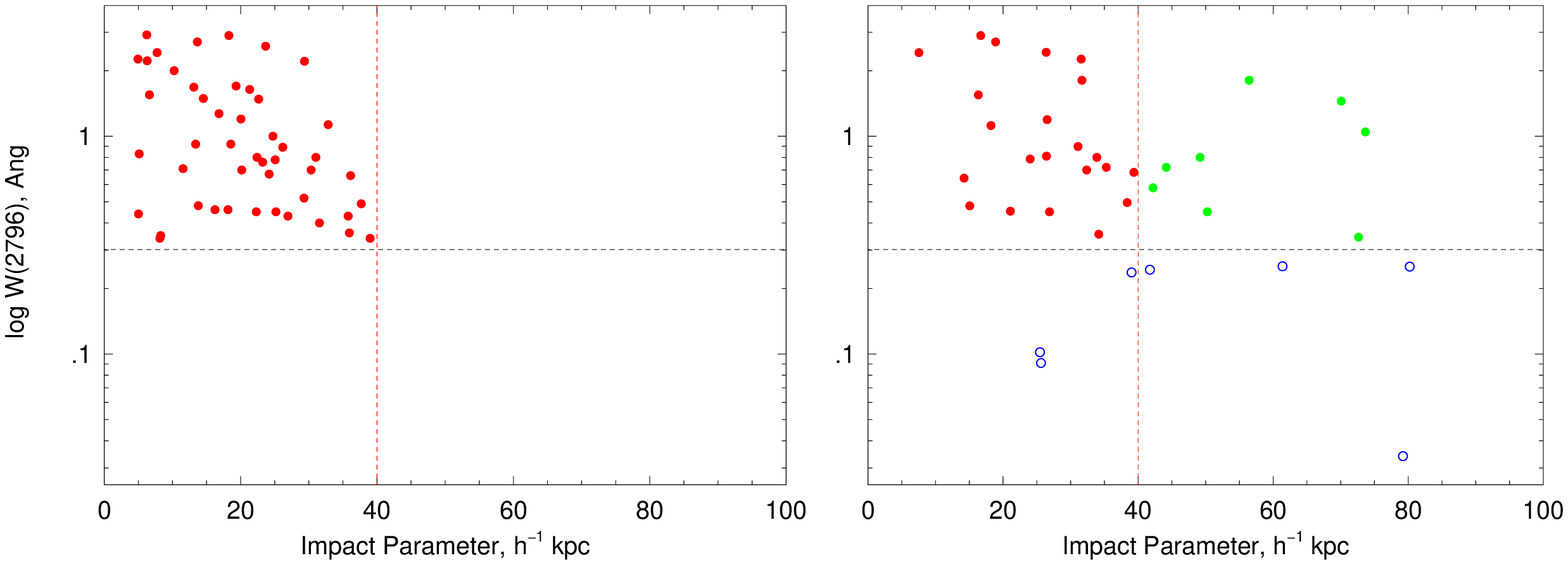}
\caption{--- The rest--frame {\cwcMgII} $\lambda 2796$ equivalent
widths versus impact parameters.  The vertical dashed line marks $D =
40~h^{-1}$, the approximate value of $R_{\ast}$, and the
horizontal dashed line is at $W_r = 0.3$~{\AA}, the separation between
strong and weak {\cwcMgII} absorbers. --- (left) The upper left area
is the region previously explored by Steidel, Dickinson, \& Persson
(1994).  --- (right) Our {\it HST\/} sample (to date).}
\label{cwcfig:wvsd}
\end{figure}

Those caveats having been stated, we have found some generally
illuminating results that are {\it not\/} sensitive to incompleteness
of redshifts in the quasar fields. In Figure~\ref{cwcfig:wvsd}, we
show the distribution of equivalent widths versus galaxy impact
parameter.  In the left panel, we show the original SDP94 distribution
for $W_{r} > 0.3$~{\AA}.  To guide the eye, we have placed a vertical
line at $D=40h^{-1}$ kpc approximately at the $R_{\ast}$ value.  We
have also placed a horizontal line at $W_{r} = 0.3$~{\AA} to demarcate
``strong'' and ``weak'' {\cwcMgII} absorbers.  Note that the SDP94
survey data lie exclusively in the range $D < 40h^{-1}$ kpc and $W_{r}
> 0.3$~{\AA}\footnote{A few of the galaxies published in the SDP94
paper have $W_{r} > 0.2$~{\AA}; however, these systems were found {\it
after the fact\/}, in that a potential absorbing galaxy was observed
in the quasar field and then {\cwcMgII} absorption was sought in
follow--up, deeper quasar spectra.}.

In the right panel of Figure~\ref{cwcfig:wvsd}, we show our {\it
HST\/} sample (to date).  Our sample includes 38 absorption systems.
In cases where we have HIRES and UVES spectra, the equivalent widths
have been remeasured.  We have also remeasured the galaxy--quasar
impact parameters in the {\it HST\/} images.  Though there is
substantial overlap between our sample and the SDP94 sample, the new
equivalent widths and impact parameters do not always exactly match
those measured in the lower resolution SDP94 data; this is simply a
resolution effect.

As can be seen, we have found galaxies associated with weak absorbers
over a wide range of impact parameters and galaxies with strong
absorbers well beyond the $R_{\ast}$ prediction.  The absorbers in the
range $70 \leq D \leq 80~h^{-1}$ kpc are particularly remarkable in
that a wide range of equivalent widths are observed.  At face value,
it would appear that there are departures from our current
conventional wisdom.  The weak systems at $D \leq 40~h^{-1}$ kpc
immediately suggest that the extended gaseous regions surrounding
galaxies are patchier than previously reported.  Moreover, the strong
{\cwcMgII} absorbers at $D \geq 40~h^{-1}$ kpc suggest that the
absorbing gas geometry is less regular than previously reported.  In
\S~\ref{cwcsec:departures}, we will highlight a few cases to
illustrate that there are clear exceptions to our current
expectations.

Since {\cwcMgII} equivalent width is strongly proportional to the
velocity spread and number of components, or ``clouds''
(\cite[Petitjean \& Bergeron 1990; Churchill {\etal}
2003]{cwcref:pb90,cwcref:cvc03}), these data suggest that the
kinematic conditions of the absorbing gas is not strongly connected to
the projected separation from the galaxy (\cite[also see Churchill
{\etal} 1996]{cwcref:csv96}).

\subsection{Morphologies, Orientations, and Gas Kinematics}

\begin{figure}[th]
\includegraphics[width=5.3in,angle=0]{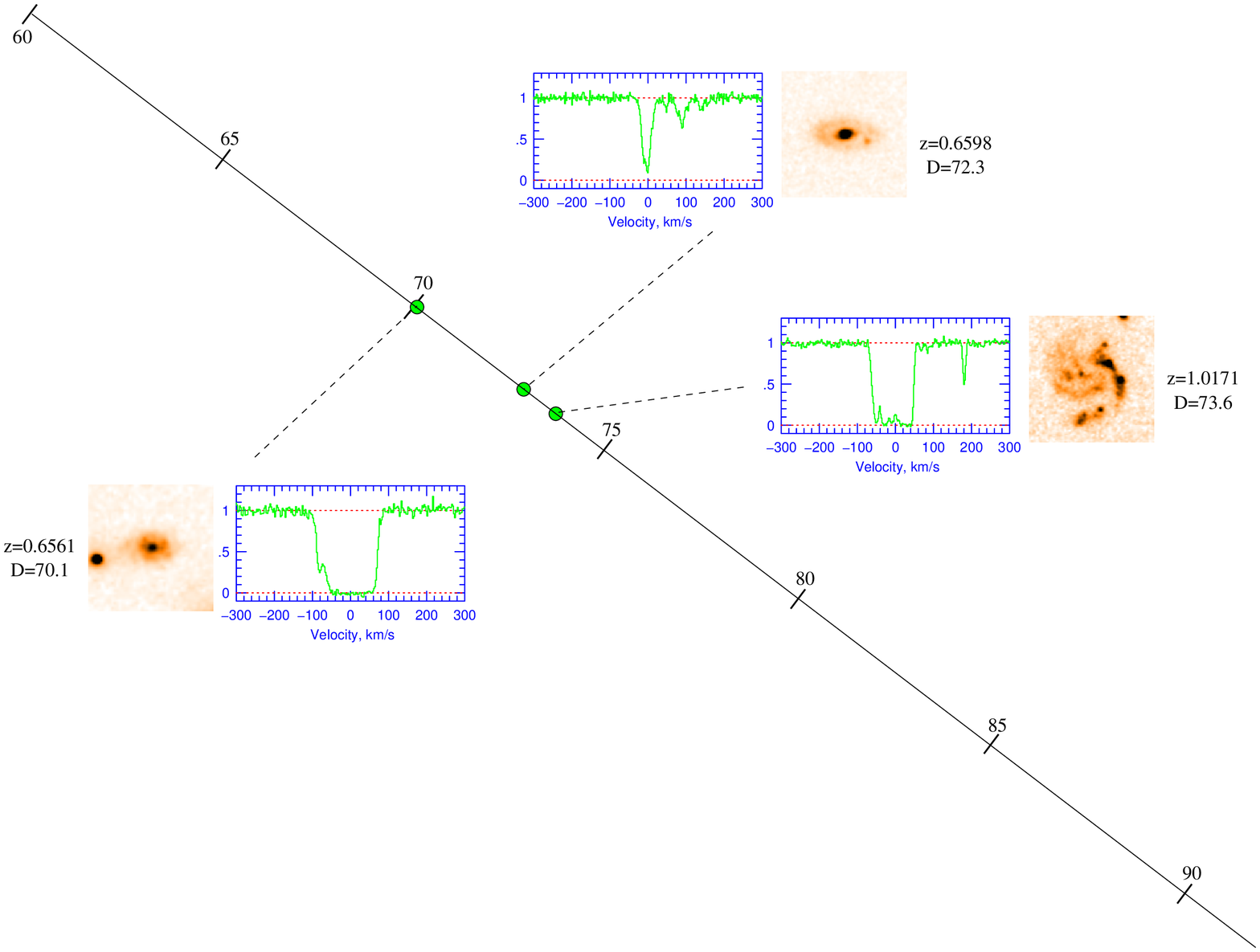}
\caption{{\it HST\/} images and HIRES/UVES {\cwcMgII} $\lambda 2796$
profile of absorbing galaxies displayed in order of increasing impact
parameter, $D$, for the range $0\leq D h \leq 90~h^{-1}$ kpc.  For the
small impact parameter systems, the diffraction spikes from the quasar
can be seen.  The axis running from upper left serves as an ``impact
parameter ruler''.  The images are $5^{\prime\prime} {\sf
x}~5^{\prime\prime}$ and oriented such that the quasar is downward in
order to illustrate the galaxy orientation with respect to the quasar
line of sight.  The absorption profiles are shown in rest--frame
velocity space over an interval of $\Delta v =600$ km~s$^{-1}$, where
the zero point is arbitrarily defined.}
\label{cwcfig:impacstr}
\end{figure}

In Figure~\ref{cwcfig:impacstr}, we present our current sample having
$W_r > 0.3$~{\AA}.  For each galaxy, the {\it HST\/} image of the
absorbing galaxy is $5^{\prime\prime} {\sf x}~5^{\prime\prime}$ and
oriented such that the quasar is downward.  This orientation provides
a consistent galaxy--quasar line of sight orientation for all galaxies
for purposes of illustration.  The HIRES and UVES absorption profiles
are of the {\cwcMgII} $\lambda 2796$ transition only, shown in
rest--frame velocity space over an interval of $\Delta v = 600$
km~s$^{-1}$.  The velocity zero point is unrelated to the galaxy
redshift; for each profile, the zero point is the optical depth median
of the absorption.  The galaxies are presented in order of increasing
impact parameter.  An ``impact parameter ruler'' runs from the upper
left to the lower right of each panel (first) $0\leq D \leq 30~h^{-1}$
kpc, (second) $30\leq D \leq 60~h^{-1}$ kpc, and (third) $60\leq D
\leq 90~h^{-1}$ kpc.

\begin{figure}[p]
\includegraphics[width=5.3in,angle=0]{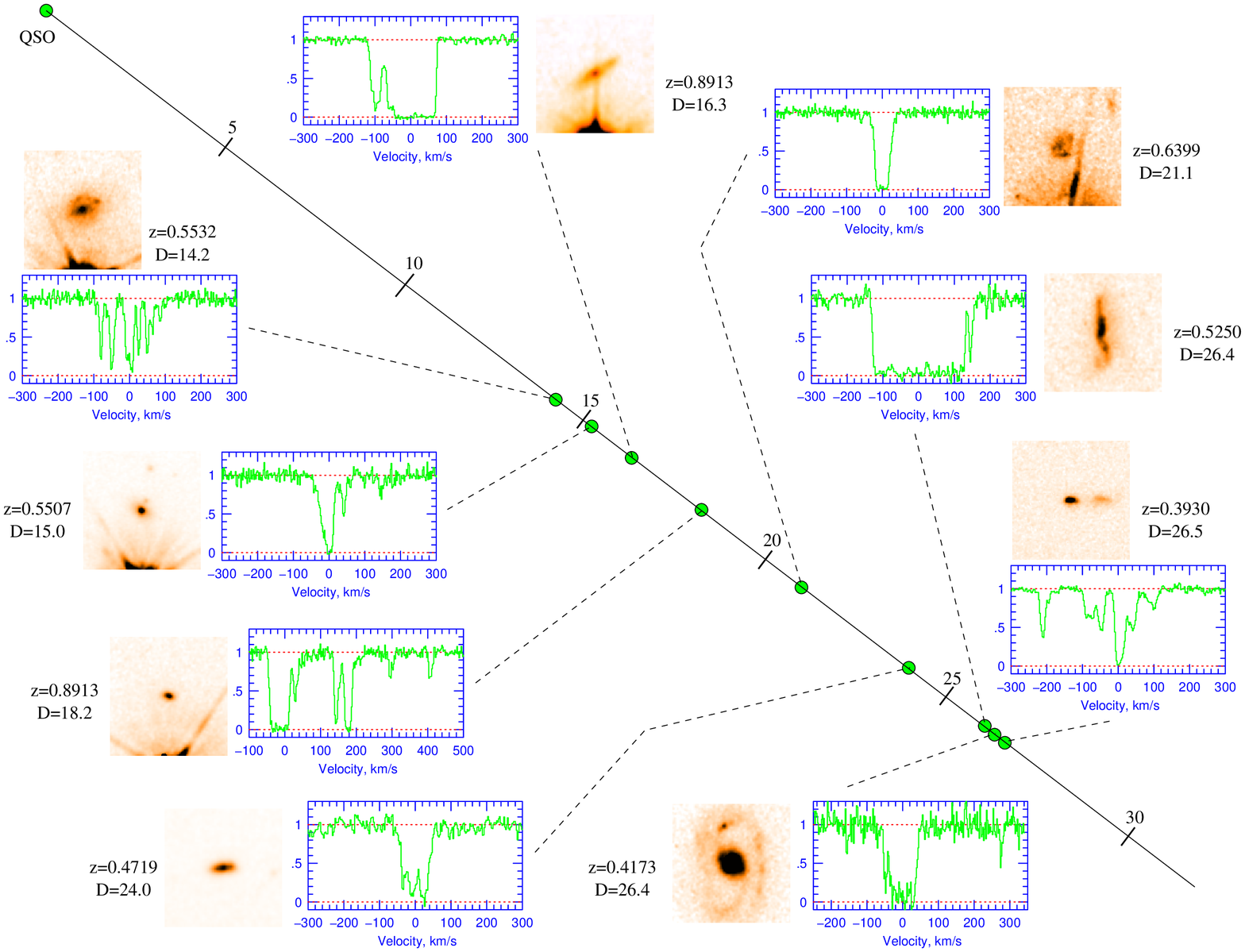}
\includegraphics[width=5.3in,angle=0]{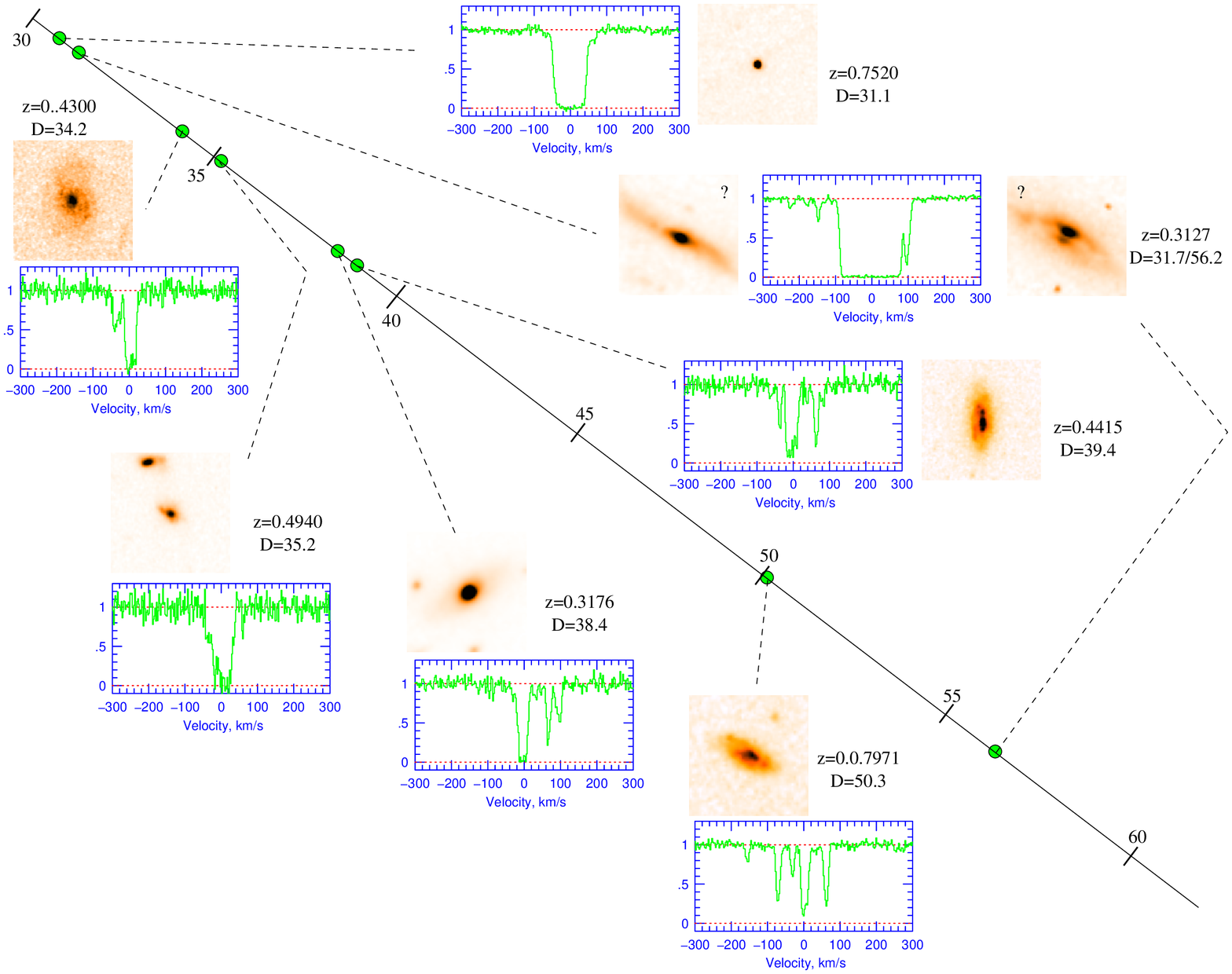}
\end{figure}

Based upon a visual examination, we can see that {\cwcMgII} absorption
selected galaxies are, as a whole, fairly normal looking.  Most
galaxies are spiral galaxies.  However, closer inspection reveals that
the majority of the galaxies have minor perturbations, and in some
cases what appears to be bright {\cwcHII} regions and/or minor
companions.  There is no trend between galaxy morphological type with
impact parameter. Inspection of the {\cwcMgII} absorption reveals a
wide variety of kinematics.  Interestingly, there are
DLA/{\cwcHI}--rich absorbers at both small and very large impact
parameters.

The galaxy orientation provides the light path geometry through the
galaxy, where the impact parameter is the closest approach to the
galactic center.  By orientation, we mean the combined projection of
the galaxy due to the galaxy inclination, $i$, and the position angle,
$\phi$, of the galaxy.  We define the position angle as the primary
angle between the galaxy major axis and a line connecting the galaxy
center and the quasar.  As a reminder, all galaxies are oriented such
that the quasar is located downward in Figure~\ref{cwcfig:impacstr}.

We have used GIM2D (\cite[Simard {\etal} 2002]{cwcref:simard02}) to
model the galaxy morphologies, and from these models measure the
galaxy position angles and inclinations (\cite[details in Kacprzak
{\etal} 2005, {\it this volume}]{cwcref:kcs05}).  These are
two--component (bulge and disk), axis--symmetric, smooth models.  As
such, perturbations in the galaxy morphologies are clearly present in
the residuals and can be quantified.  We find significant departures
from smooth morphologies in {\cwcMgII} absorbing galaxies that would
otherwise be missed without modeling.

We have searched for correlations between galaxy orientation
parameters and {\cwcMgII} absorption properties.  Galaxy orientation
parameters (such as $\cos \phi $, $\sin \phi$, $\cos i $, $\sin i$,
$\cos \phi \cos i$, $\sin \phi \cos i$, etc.), do not correlate with
absorption properties.  This might be interpreted to mean that the
halos are spherical; however, it can just as well be interpreted to
mean that the absorbing gas can be extended in any orientation with
respect to the galaxy.  For the latter scenario to be favored, we
would expect very weak to non--absorption at moderate impact
parameters in some fraction of the galaxies.  We will show below that
this is precisely what we are finding.  Interestingly, we find a
$3.2~\sigma$ correlation between the {\cwcMgII} $\lambda 2796$
equivalent width and the galaxy morphological ``asymmetries''
normalized by the impact parameter.  The details and interpretations
are discussed in Kacprzak {\etal} ({\it this volume}).

\subsection{Weak Systems and Galaxies}

In our sample, there are seven weak {\cwcMgII} systems for which we
have {\it HST\/} images of the quasar fields.  Follow up spectroscopy
has revealed, in each case, that weak systems are associated with
galaxies.  Furthermore, these galaxies appear to have fairly normal
morphologies, if somewhat sub--$L^{\ast}$ luminosities.  This suggests
that weak {\cwcMgII} absorbers do not select a particular class of
galaxy (i.e., LSB, dwarf) or post--star forming object.

\begin{figure}[h]
\includegraphics[width=5.3in,angle=0]{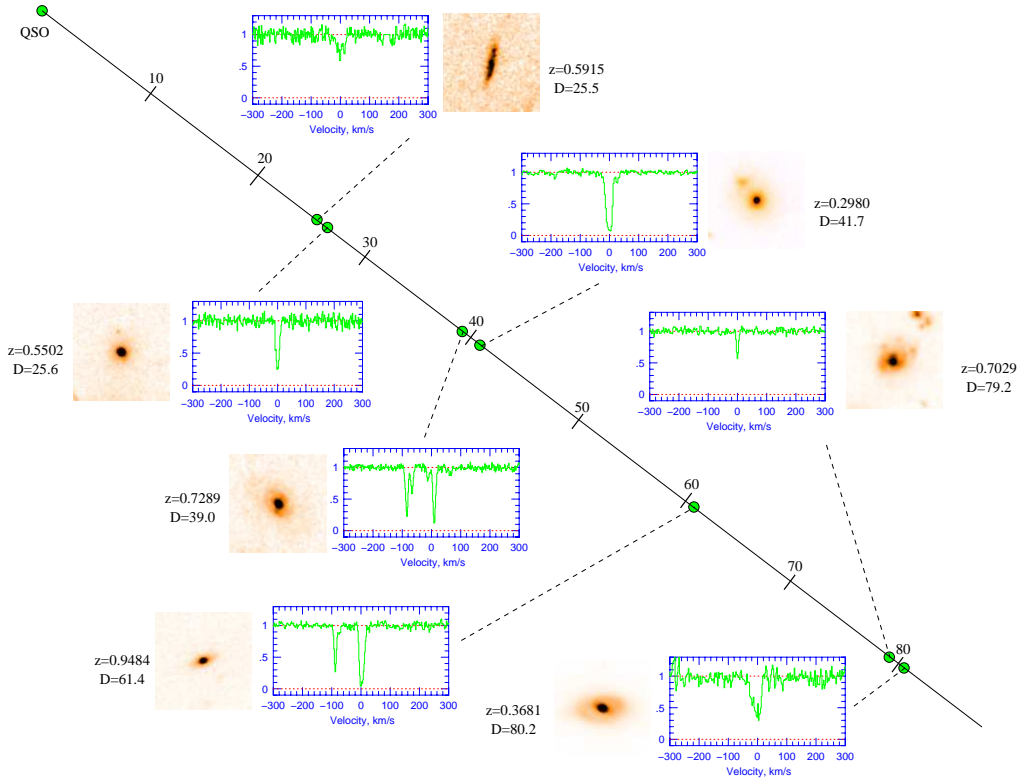}
\caption{Same as Figure~\ref{cwcfig:impacstr} but for the the weak
{\cwcMgII} absorbers (defined to have $W_{r} < 0.3$~{\AA}) over the
range $0\leq D \leq 80~h^{-1}$ kpc.  We have found seven weak
{\cwcMgII} absorbers clearly associated with galaxies.}
\label{cwcfig:impacwk}
\end{figure}

Two of the seven galaxies are at impact parameters $D~h^{-1} \simeq
80$ kpc.  On the other hand, three (almost four) of the seven galaxies
are at $D~h^{-1} \leq 40$ kpc, and would have been labeled as
non--absorbing galaxies in the SDP94 survey.  In fact, in two cases,
the galaxy at the weak {\cwcMgII} absorption redshift was
misidentified to be associated with strong {\cwcMgII} absorption at a
different redshift due to the lack of spectroscopic redshifts.  Such
reidentifications imply that the covering factor for $W_r > 0.3$~{\AA}
is probably less than unity (\cite[see Charlton \& Churchill
1996]{cwcref:cc96}).

We cannot yet constrain the distribution of impact parameters for weak
systems; to do so would require a larger sample.  However, if we
consider that we have observed two absorbers at $D~h^{-1} \geq 70$ kpc
that classify as DLA/{\cwcHI}--rich and three weak absorbers at
$D~h^{-1} \leq 40$ kpc, we would probably be incorrect to assume that
weak {\cwcMgII} absorbers are preferentially found at large impact
parameters and stronger absorbers are found only at smaller impact
parameters.  As such, we may be finding that weak {\cwcMgII}
absorption arises in galaxies with a distribution of impact parameters
similar to that of strong {\cwcMgII} absorbers.  For this to be the
case, the $N({\cwcHI})$ around the galaxies must range over several
orders of magnitude at all impact parameters (probably becoming more
varied with increasing impact parameter).

\subsection{More to the Story?}
\label{cwcsec:departures}

\begin{figure}[h]
\includegraphics[width=5.1in,angle=0]{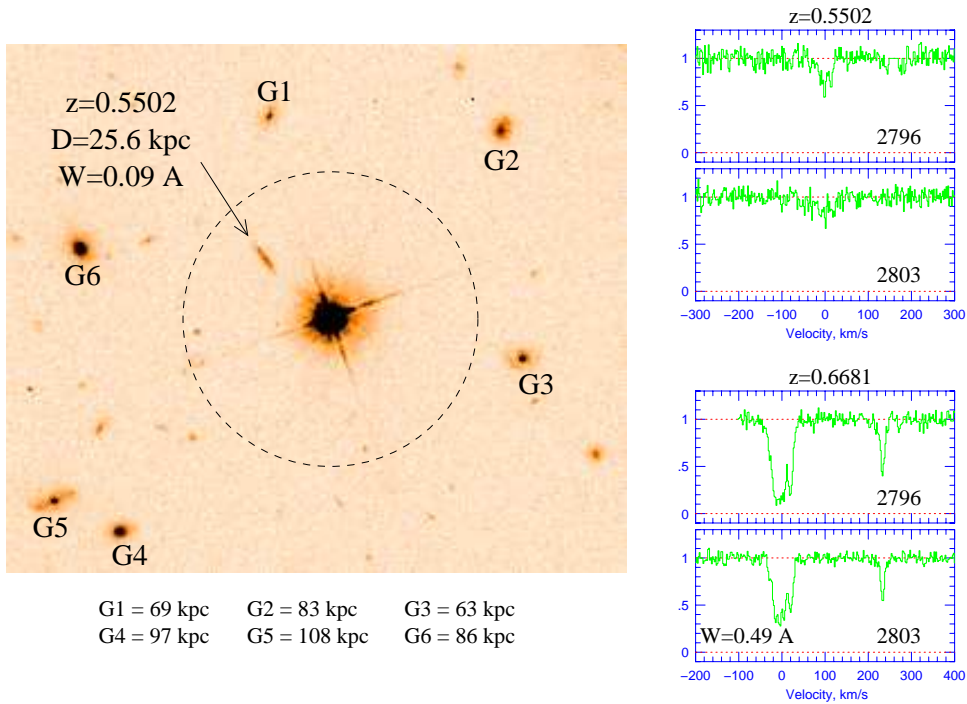}
\caption{--- (left) The WFPC-2 {\it HST\/} image of the Q$1222+228$
field.  The $z=0.5502$ absorbing galaxy is identified; it is a weak
{\cwcMgII} absorber.  There is strong {\cwcMgII} absorption at
$z=0.6681$, but the galaxy remains unidentified.  Absorber candidates
are labeled G1 through G6, but they all lie outside of the $R_{\ast}$
prediction shown as the dashed circle centered on the quasar.  ---
(right) The HIRES absorption profiles for the $z=0.5502$ and
$z=0.6681$ systems. The absorption profiles are shown in rest--frame
velocity space over an interval of $\Delta v =600$ km~s$^{-1}$, where
the zero point is arbitrarily defined.}
\label{cwcfig:q1222}
\end{figure}

\begin{figure}[h]
\includegraphics[width=5.1in,angle=0]{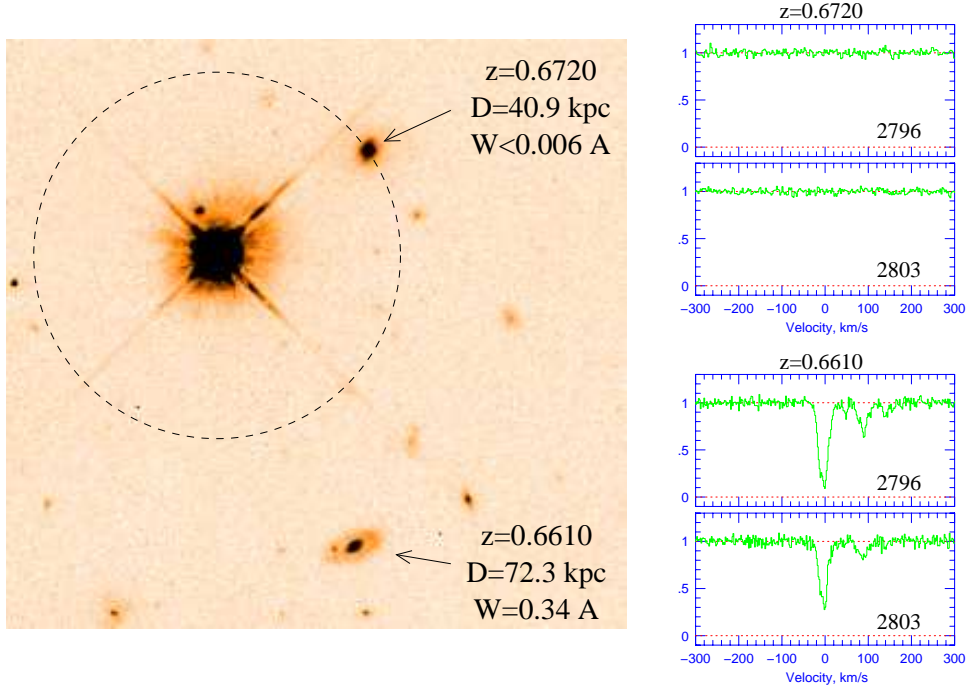}
\caption{--- (left) The WFPC-2 {\it HST\/} image of the Q$1317+274$
field.  The $z=0.6610$ absorbing galaxy is at $72~h^{-1}$ kpc, well
outside the $R_{\ast}$ prediction, shown as a dashed circle centered
on the quasar.  A second galaxy at $41~h^{-1}$ kpc, is confirmed to
have $z=0.6720$. --- (right) The HIRES/Keck absorption profiles for
the $z=0.6610$ systems (lower) and the HIRES spectrum at the expected
location of the $z=0.6720$ galaxy.  There is no absorption to a
rest--frame limit of $W_r \leq 0.006$~{\AA} ($3~\sigma)$.}
\label{cwcfig:q1317}
\end{figure}

In Figures~\ref{cwcfig:q1222} and \ref{cwcfig:q1317}, we show two
quasar fields that further suggest that {\cwcMgII} absorption is
patchier than previously thought.  If the examples of these fields are
moderately common, it would require modification of our current views.

In the Q$1222+228$ field, the first absorber known was at $z=0.6681$.
The {\cwcMgII} profiles are shown in the lower right of
Figure~\ref{cwcfig:q1222}.  It is a strong absorber easily detected in
the spectrum of \cite{cwcref:ss92}.  The edge--on spiral galaxy at $D
\simeq 26~h^{-1}$ kpc, labeled in the {\it HST\/} image of
Figure~\ref{cwcfig:q1222}, was assumed to be the host of the
{\cwcMgII} absorption.  However, a LRIS/Keck~I spectrum of the galaxy
places it at $z=0.5502$, a redshift at which \cite{cwcref:weakI}
previously reported a weak {\cwcMgII} absorber.  The strong $z=0.6681$
{\cwcMgII} absorbing galaxy remains unidentified.  As labeled in
Figure~\ref{cwcfig:q1222}, there are six candidate galaxies, G1--G6,
for the $z=0.6681$ {\cwcMgII} absorber.  At this redshift, they lie in
the impact parameter range $63 \leq D \leq 108~h^{-1}$ kpc, which is
well beyond the $R_{\ast}$ expectation (shown by the dashed circle
centered on the quasar).  We are in the process of obtaining the
spectra of these six galaxies; if/when the absorber is identified, it
may yet be another example of strong absorption at large impact
parameter.

As seen in Figure~\ref{cwcfig:q1317}, a bright $z=0.6720$ galaxy with
$D \simeq 41~h^{-1}$ kpc in the Q$1317+274$ field is a classic example
of the type of galaxy expected to host strong {\cwcMgII} absorption.
However, at the expected location of the {\cwcMgII} doublet in the
HIRES spectrum, we find no absorption to a $3~\sigma$ limit of $W_r <
0.006$~{\AA} (upper right panel).  This limit is a factor of three
below current surveys for weak {\cwcMgII} absorbers and a factor of 50
below the expected absorption strength for this galaxy.  For
{\cwcMgII}, this ``mother of all non--absorbers'' provides an emphatic
accentuation to our other findings that the halos around galaxies are
quite patchy and that the covering factor for $W_r>0.3$~{\AA} and $D
\leq 40~h^{-1}$~kpc must be less than unity.

In the Q$1317+274$ field, there is {\cwcMgII} absorption at $z=0.6610$
with $W_r = 0.34$~{\AA}.  The profiles are shown in the lower right of
Figure~\ref{cwcfig:q1317}.  The absorption is associated with a bright
galaxy at $D \simeq 72~h^{-1}$ (\cite[see Steidel {\etal}
2002]{cwcref:steidel02}).  This latter galaxy would not be expected to
give rise to absorption with this equivalent width.  This example
suggests that the geometry of galaxy ``halos'' is not necessarily
spherical, but can apparently be highly extended.

In our sample, there are two absorbers that arise in ``double''
galaxies (see Figure~\ref{cwcfig:impacstr}).  The first is at redshift
$z=0.4940$ with an approximate impact parameter of approximately $D =
35.2~h^{-1}$ kpc (galaxy nearest the quasar).  These galaxies appear
to be in the process of merging, for there is a tidal tail attached to
each galaxy that arcs back toward its companion.  Note however, that
the absorption does not have the tall--tale signs of a ``double''
{\cwcMgII} absorber (\cite[Churchill {\etal} 2000b]{cwcref:archiveII})
based upon the less than 200 km~s$^{-1}$ velocity spread and absence
of high velocity kinematic subsystems.

The second pair is at $z=0.3127$.  One is at $D = 35.2~h^{-1}$ kpc and
the second is at $D = 56.2~h^{-1}$ kpc.  The projected separation
between the galaxies is only $36.5~h^{-1}$ kpc, but it is not clear if
they are gravitationally bound or part of a group.  Originally, the
closer in galaxy was assumed to be the only absorber.  This case
illustrates that additional care must be taken to acquire a full
census of the galaxies in absorber fields (\cite[also see Bowen,
Blades, \& Pettini 1995]{cwcref:bbp95}).  Interestingly, both galaxies
have almost identical orientations with respect to the quasar line of
sight.  There appears to be a warped disk in the $D = 35.2~h^{-1}$ kpc
galaxy and perhaps some disturbance and LMC/SMC--like satellite
galaxies associated with the $D = 56.2~h^{-1}$ kpc.  It may be that
one can never establish the host galaxy of the absorption for such
cases; however, obtaining the kinematics of the galaxies themselves
(i.e., rotation curves) would serve to constrain the dynamical
relationships between the galaxies and the absorption.

\section{Sorting it Out}
\label{cwcsec:discuss}

We have shown several galaxies where the impact parameters and
strengths of {\cwcMgII} absorption run counter to our current
conventional wisdom.  Our data would suggest a covering factor for
$W_r > 0.3$~{\AA} less than unity (how much less we cannot say) and a
geometry that is not necessarily characterized as spherical.

In order to firmly establish the character of {\cwcMgII} absorbing gas
connected to galaxies we must quantify the covering factor and extent
of the absorption as a function of absorption strength and galaxy
properties.  However, any survey {\it must\/} include the distribution
of very weak and non--absorbing galaxies.  Due to the sensitivity of
HIRES and UVES quasar spectra, the definition of a non--absorbing
galaxy can now be based upon equivalent width limits of $W_r \simeq
0.02$~{\AA}, a factor of 15 below previous surveys.  It is imperative
that the cataloging of galaxies in quasar fields be complete to a
limiting absolute magnitude and to a fixed projected physical
separation from the quasar.

It is not clear that the use of control fields (those in which no
{\cwcMgII} absorption is seen in the quasar spectra) is the best
approach to cataloging non--absorbing galaxies.  It is quite possible
that the majority of these quasars could lie in the direction of
sparse galaxy fields so that the galaxies would preferentially be at
large impact parameters.  Because the quasars are known {\it a
priori\/} to lack absorption, it is also possible that control fields
could preferentially select galaxies with ``special'' histories that
yield highly patchy halos or small halo geometries.  These
considerations may partially explain the lack of non--absorbing
galaxies at small impact parameters reported by
\cite{cwcref:steidel95}.  On the contrary, most quasar with {\cwcMgII}
absorption have several galaxies in their fields.  There is no reason
to suppose that, on average, a galaxy in a field with other absorbers
at different redshifts would preferentially be a non--absorbing
galaxy.

It can be argued that the whole methodology of galaxy selection should
be turned around; quasar fields should be surveyed for galaxies before
any knowledge of absorption in the quasar spectra.  However, the
galaxy morphologies are a keystone ingredient to these studies and it
would be difficult to successfully motivate wholesale {\it HST\/}
imaging of such fields.  Obtaining {\it HST\/} images of fields with
known absorbers guarantees some galaxies for galaxy--absorber
connection analysis and provides morphologies of galaxies that
classify as non--absorbers.

In the end, it will be studies which incorporate both the galaxy
kinematics and the absorption kinematics that hold the greatest
promise for promoting our understanding of the nature of extended
gaseous regions associated with galaxies.

\section{Conclusions}
\label{cwcsec:concl}

We have presented a brief history and review of {\cwcMgII} absorption
selected galaxies.  In addition, we have presented new results and
inferences regarding the nature of {\cwcMgII} absorbing gas in the
vicinity of galaxies.  Our data include high spatial resolution {\it
HST\/} images of the quasar fields and HIRES and UVES high velocity
resolution spectra of the absorbing gas.  We have carefully confirmed
the redshifts of all galaxies in our sample to be coincident with the
absorption redshifts.  Our sample, to date, comprises 38 {\it HST\/}
imaged {\cwcMgII} absorbing galaxies, 26 of which we also have HIRES
and/or UVES spectra of the {\cwcMgII} absorption.  We are currently
working to obtain the HIRES and UVES spectra for the remaining 12
absorbers.  We summarize our main points below:

\begin{enumerate}

\item The extended gaseous envelopes surrounding intermediate redshift
galaxies are more patchy than previously reported.  Our survey data
are suggestive that the {\cwcMgII} absorbing galaxy ``halos'' have a
covering factor less than unity and geometries that depart from
spherical.  A complete and unbiased survey of quasar fields using {\it
HST\/} images and high--resolution quasar spectra are required to
quantify these statements.

\item We have made first steps toward quantifying the morphologies of
galaxies hosting {\cwcMgII} absorption.  From the galaxy orientations,
we have constrained the quasar sightline paths through these galaxies.
We have searched for correlations with {\cwcMgII} absorption
properties, including velocity spread, velocity asymmetry, number of
``clouds'', equivalent width, doublet ratio, and total column density.
There are no significant correlations between galaxy orientation and
gas kinematics, suggesting that the halo geometries and gas velocity
fields are, in general, not strongly linked to those of the galaxies

\item Some (and possibly the majority) of weak {\cwcMgII} absorbers
are associated with normal galaxies over a range of impact parameters.
We have found direct association in seven of seven cases for which we
have obtained the spectroscopic redshifts of the candidate galaxies.
This suggests that weak {\cwcMgII} absorbers do not select a
particular class of galaxy (i.e., LSBs, dwarfs, etc.) or post--star
forming object.

\end{enumerate}

We have emphasized the importance of a complete survey of galaxies in
quasar fields to a fixed absolute magnitude and fixed galaxy--quasar
separation.  It is vital that we know the redshifts of both absorbing
and non--absorbing galaxies in the fields in order to quantify the
covering factor and constrain the geometry of the absorbing gas.
Since our absorption line spectra are sensitive to $W_r \simeq
0.02$~{\AA} (and even better in some cases), we will be able to
examine these issues to a higher level of sensitivity than previous
studies.  In our survey, a non--absorbing galaxy will be one where the
the limits on {\cwcMgII} absorption is roughly a factor of ten more
sensitive than the previous $W_r < 0.3$~{\AA} criterion.  Such a
survey of the quasar fields will be very time intensive and will
require systematic and long--term effort.  We are exploring the
possibility of using photometric redshifts to place first--order
constraints on the galaxies in these fields.

What we wish to emphasize even more strongly is the need to obtain
high quality spectra of the galaxies known to host {\cwcMgII}
absorption.  In particular, it is vital that we obtain accurate
redshifts (with uncertainties less than $\simeq 20$~km~s$^{-1}$).
This will allow us to constrain the relative velocities of the gas and
the galaxy.  Most desirable are rotation curves for the galaxies in
order to extend the studies of \cite{cwcref:steidel02} and
\cite{cwcref:ellison03}.  We now have a reasonably sized sample and
can examine the kinematic connections between galaxies and {\cwcMgII}
absorption for a wide variety of galaxy orientations and galaxy
kinematics (once rotations curves are obtained).

It is a detailed case--by--case study of the galaxy kinematics --
absorbing gas kinematics connections that hold the greatest promise
for constraining the role of gas in galaxy evolution.  These are the
data that will ultimately provide the most meaningful constraints on
galaxy--galaxy halo models (which will soon incorporate mock spectra
in a manner similar to cosmological simulations).  We can constrain
the line--of--sight light path through the galaxies; however, only by
documenting the velocities of the gas relative to those of the
galaxies can we compare the observed conditions in early epoch
galaxies to models of galactic fountains, superbubbles, galactic
winds, high velocity clouds, and accretion via minor mergers.  This is
how we will finally understand the nature of extended gaseous
envelopes around normal galaxies.

\begin{acknowledgments}
We would like to acknowledge partial support from NASA, the NSF, and
the IAU.  Naoto Kobayashi, Michael Murphy, Michael Rauch, Wal Sargent,
and Alice Shapley all made important contributions to the new work
presented herein.
\end{acknowledgments}


\end{document}